\documentclass[12pt,epsfig]{article}
\newcommand{\ba}{\begin{array}{c}}
\newcommand{\baz}{\begin{array}{cc}}
\newcommand{\bad}{\begin{array}{ccc}}
\newcommand{\bav}{\begin{array}{cccc}}
\newcommand{\ea}{\end{array}}

\usepackage{amssymb} 
\usepackage{epsfig}
\usepackage{float}
\newcommand{\be}{\begin{equation}}
\newcommand{\ee}{\end{equation}}
\newcommand{\bea}{\begin{eqnarray}}
\newcommand{\eea}{\end{eqnarray}}
\begin{document} 

\begin{flushright}
SISSA 34/2004/EP\\
hep-ph/0405237
\end{flushright}

\begin{center}
\bf{Nuclear Matrix Elements of $0\nu\beta\beta$-Decay:
Possible Test of the Calculations}
\end{center}

\begin{center}
S. M. Bilenky 
\end{center}

\begin{center}
{\em  Joint Institute
for Nuclear Research, Dubna, R-141980, Russia, and\\
Scuola Internazionale Superiore di Studi Avanzati, 
I-34014 Trieste, Italy.}
\end{center}

\begin{center}
S.T. Petcov 
\footnote{Also at: Institute of Nuclear Research and
Nuclear Energy, Bulgarian Academy of Sciences, 
1784 Sofia, Bulgaria.}
\end{center}

\begin{center}
{\em Scuola Internazionale Superiore di Studi Avanzati, and\\
INFN-sezione di Trieste, I-34014 Trieste, Italy}
\end{center}

\begin{abstract}
A possible model independent test of the 
theoretically calculated nuclear matrix elements 
of $0\nu\beta\beta$-decay is proposed. 
The test can be accomplished if 
$0\nu\beta\beta$-decay 
of three (or more) nuclei is observed.
The selection of the nuclei for the next generation
of $0\nu\beta\beta$-decay experiments should
be done taking into account 
considerations regarding the possibility
to test the nuclear matrix element calculations.
The test proposed allows also to check the dominance of the   
Majorana mass mechanism of violation 
of the total lepton charge.

\end{abstract}

\section{Introduction}

The status of the problem of  
neutrino mixing changed drastically during the last 
several years: in the Super-Kamiokande (SK) atmospheric neutrino 
\cite{SK}, SNO solar neutrino 
\cite{SNO,SNOsalt}
and KamLAND reactor antineutrino 
\cite{Kamland} experiments  {\em model independent 
evidences of neutrino oscillations} were obtained.
All neutrino oscillation data, except,
the data of the LSND experiment \cite{LSND}~ 
\footnote{In the
accelerator LSND experiment indications 
in favor of the transitions
$\bar \nu_{\mu}\to\bar \nu_{e}$  with $(\Delta m^{2})_{\rm{LSND}}\simeq 
1\rm{eV}^{2}$ were obtained. 
The  LSND results are being tested 
in the MiniBooNE experiment
at Fermilab \cite{MiniB}.},
can be described if we assume the existence of
three-neutrino mixing in vacuum:
\be
\nu_{{l}L}(x) = \sum^{3}_{i=1}U_{{l}i} \, \nu_{iL}(x).
\label{1}
\ee
Here  $\nu_{i}(x)$ is the field on neutrino with mass $m_{i}$
and $U$ is the unitary PMNS \cite{BP,MNS} mixing matrix.

   The SK atmospheric neutrino data are  best  
described in terms of two-neutrino 
$\nu_{\mu} \to \nu_{\tau}$ oscillations.
From the analysis of the data 
the following best-fit values of the oscillation 
parameters were found \cite{SK}:
\be
|\Delta m^{2}_{32}|=2\cdot 10^{-3}\rm{ eV}^{2},\,
~~~\sin^{2}2 \theta_{23}=1.0 \,~~~~
(\chi^{2}_{\rm{min}}= 170.8/ 170\,\rm{d.o.f.}). 
\label{2} 
\ee
At the  90\% C.L. one has:
\be
1.3 \cdot 10 ^{-3}\leq |\Delta m^{2}_{32}| \leq 3.0 
\cdot 10 ^{-3}\rm{eV}^{2}, \,~~~~~\sin^{2}2 \theta_{23} >0.9 .
\label{3}
\ee

  The results of all solar neutrino experiments 
can be explained by $\nu_{e}\to \nu_{\mu,\tau}$ 
transitions in matter. In the KamLAND experiment,
$\bar\nu_{e} $ disappearance due to transitions 
$\bar{\nu}_{e}\to \bar{\nu}_{\mu,\tau}$ in vacuum 
was observed. From a global two-neutrino oscillation 
analysis of the solar and KamLAND data 
(performed under the assumption 
of CPT-invariance),
the following best-fit values 
of the relevant oscillation parameters
were obtained \cite{SNOsalt}:
\be
\Delta m^{2}_{21} = 7.1\cdot 10^{-5}\rm{eV}^{2},
\,~~~~~\tan^{2}\theta_{12}=0.41.
\label{4}
\ee
\noindent In a similar 3-neutrino oscillation 
analysis of the solar neutrino, 
KamLAND and CHOOZ \cite{CHOOZ} data,
performed in \cite{Sand}, it was found  
that at 90\% C.L. one has:
\be
\ba
5.6\cdot 10^{-5}\leq  \Delta m^{2}_{21}
\leq 9.2\cdot 10^{-5}\rm{eV}^{2},\,~~
 0.23 \leq \sin^{2}\theta_{12}\leq 0.38, 
~~{\rm for}~~ \sin^2 \theta_{13} = 0.0 , \\
6.1\cdot 10^{-5}\leq  \Delta m^{2}_{21}
\leq 8.5\cdot 10^{-5}\rm{eV}^{2},\,~~
 0.25 \leq \sin^{2}\theta_{12}\leq 0.36, 
~~{\rm for}~~ \sin^2 \theta_{13} = 0.04,
\ea
\label{5}
\ee
\noindent where $\theta_{13}$ is the mixing 
angle limited by the reactor CHOOZ and Palo Verde 
experiments \cite{CHOOZ,PVerde}.
The negative results of the 
CHOOZ \cite{CHOOZ} and Palo Verde \cite{PVerde} 
experiments are very 
important for understanding the 
pattern of neutrino mixing and oscillations.
In these experiments no disappearance 
of $\bar \nu_{e}$ was 
observed. From the 90\% C.L.
exclusion curve obtained from 
the analysis of the data of the CHOOZ experiment, the 
following bound can be derived
\be
\sin^{2}\,\theta_{13} < 5\cdot 10^{-2}. 
\label{6}
\ee
%
\noindent The same result was obtained in 
\cite{Sand} in a global 3-neutrino oscillation 
analysis  of the solar, KamLAND and CHOOZ data 
with $|\Delta m^{2}_{32}|$ taken to lie 
in the interval eq. (\ref{3}).

  There are two general 
theoretical possibilities 
for the fields of neutrinos 
with definite masses $\nu_{i}(x)$ 
(see, e.g., \cite{BiPet87}):
\begin{enumerate}
\item
If the total lepton charge
$L = L_{e} + L_{\mu} +L_{\tau}$ is conserved,  
$\nu_{i}(x)$ are {\em Dirac fields} of 
neutrinos $\nu_{i}$ (L=1) and antineutrinos 
$\tilde{\nu}_{i}$ (L=-1).
\item
If there are no conserved lepton charges, $\nu_{i}(x)$ 
are {\em Majorana fields} 
which satisfy the condition 
\be
\nu^{c}_{i}(x)= C\,\bar\nu^{T}_{i}(x)=\nu_{i}(x),
\label{7}
\ee
$C$ being the charge conjugation matrix, and 
neutrinos $\nu_{i}$ are {\it Majorana particles}. 
\end{enumerate}

  The solution of the problem of the nature 
of the neutrinos with definite mass - Dirac 
or Majorana, will  
be of fundamental importance 
for the understanding of the 
origin of small neutrino masses 
and of the pattern of neutrino mixing.  

   The investigation of neutrino oscillations 
does not permit to determine the nature of 
massive neutrinos \cite{BHP80,Lang87}. Processes 
in which the total lepton charge 
$L$ is not conserved and 
changes by two units,
must be studied for that purpose.
These processes are allowed if the 
massive neutrinos are Majorana particles.
Experiments searching for
neutrinoless double 
$\beta $- decay of even-even nuclei
(see, e.g., \cite{Doi85,Faessler,Suhonen,Vergados,Elliot,Morales02,BGGM}),
\be
(\rm{A,Z}) \to (\rm{A,Z +2}) +e^- +e^-.
\label{8}
\ee
have the highest sensitivity to the 
nonconservation of the total lepton 
charge and to Majorana neutrino 
masses. The matrix element of $0 \nu \beta \beta$ -decay is 
proportional to the effective Majorana mass 
(see, e.g., \cite{BiPet87,BPP1}):
\begin{equation}
|m_{ee}| =  \left| m_1 |U_{\mathrm{e} 1}|^2 
+ m_2 |U_{\mathrm{e} 2}|^2~e^{i\alpha_{21}}
 + m_3 |U_{\mathrm{e} 3}|^2~e^{i\alpha_{31}} \right|~,
\label{9}
\end{equation}
\noindent where 
$U_{ej}$, $j=1,2,3$, are the elements of the 
first row of the PMNS
neutrino mixing matrix $U$,
$m_j > 0$ is the mass of the Majorana neutrino $\nu_j$,
and $\alpha_{21}$ and $\alpha_{31}$ 
are two Majorana CP-violating phases
\cite{BHP80,Doi81}.
One can express 
\cite{SPAS94} (see also, e.g., \cite{BGGKP99,BPP1})
the two heavier neutrino masses 
and the elements $|U_{ej}|$
in $ |m_{ee}|$
in terms of the lightest 
neutrino mass,
$\Delta m^2_{21}$,
$\Delta m^2_{32}$, 
and of $\theta_{12}$ and $\theta_{13}$, respectively.

   The results of a large number 
of experiments searching 
for $0 \nu \beta \beta$ -decay
are available at present 
(see, e.g., \cite{Morales02,Zdesenko,Gratta,PDG}).        
In the Heidelberg-Moscow experiment \cite{HM}
the most stringent lower bound on the half-life
of $0\,\nu \beta\,\beta $-decay of $^{76}\rm{Ge }$ 
has been obtained
\footnote{Indications for 
$0\,\nu \beta\,\beta $-decay of $^{76}\rm{Ge }$
with a rate corresponding to
$0.11 \ {\rm eV} \leq  
|m_{ee}|  \leq  0.56$ eV
(95\% C.L.), are claimed to 
have been obtained in \cite{Klap01}. The
results announced in \cite{Klap01} have been 
criticized in \cite{bb0nu02}. Even stronger evidence 
has been reported recently in \cite{Klap04}. These claims
will be checked in the currently running and future
$0\,\nu \beta\,\beta $-decay experiments. 
However, it may take a very long time 
before comprehensive checks could be completed.
}:
\be
T^{0\nu}_{1/2} > 1.9 \cdot 10^{25}\,{\rm y}\,~~ (90\% \,\rm{C.L.}).
\label{10}
\ee
In the cryogenic experiment CUORICINO  
the following lower bound on 
the $0\,\nu \beta\,\beta $-decay half-life of 
$^{130} \rm{Te }$ was  recently  
reported \cite{Cuoricino}:
\be
T^{0\nu}_{1/2} > 7.5 \cdot 10^{23}\, \rm{y}
\label{11}
\ee
  
\noindent 
 Taking into account the result of different 
calculations of the relevant nuclear matrix elements,
the following upper bounds
on the effective Majorana mass $|m_{ee}|$ 
can be inferred from the limits (\ref{10}) and (\ref{11}):
\be
|m_{ee}| < (0.3-1.2)\,~\rm{eV},~~~ \quad|m_{ee}| < (0.3-1.7)\,~\rm{eV}.
\label{12}
\ee

\noindent The NEMO3 experiment searching for 
$0\,\nu \beta\,\beta $-decay of a number of different
nuclei ($^{100}\rm{Mo }$, $^{82} \rm{Se}$, etc.)
and aiming at a precision of 
$|m_{ee}| \sim 0.1\,~\rm{eV}$, is 
successfully taking data at present \cite{NEMO3}.

   Many new projects of experiments 
searching for neutrinoless 
double $\beta$-decay of  $^{76}$\,Ge, $^{136}$\,Xe, $^{130}$\,Te,  
$^{100}$\,Mo and other nuclei, 
are under research and 
development at present 
(see, e.g., \cite{Elliot,Morales02,Gratta,Avignone} ).
The goal of the future experiments is to 
reach a sensitivity 
\be
|m_{ee}| \simeq \rm{\,~few} \times 10^{-2}~\rm{eV}.
\label{13}
\ee

    In order to obtain information about 
the effective Majorana mass $|m_{ee}|$
from the results of 
the $0 \nu \beta \beta$-decay experiments, 
the corresponding $0 \nu \beta \beta$-decay 
nuclear matrix elements must be known.
A large number of calculations of 
the nuclear matrix elements 
of the $0 \nu \beta \beta$-decay 
exist in the literature 
(see, e.g., \cite{Faessler,Suhonen,Elliot}). 
The results of different calculations differ 
by a factor three or more. We will propose here a 
possible test of the nuclear matrix elements calculations. 
It can be accomplished if 
$0 \nu \beta \beta$-decay of {\em several nuclei} 
will be observed in the future experiments.


\section{The effective Majorana mass}


  The neutrino oscillation data allow to predict 
the possible ranges of values of the 
effective Majorana mass. 
The prediction depend strongly on the character of 
the neutrino mass spectrum and 
on the value of the lightest neutrino mass
(see, e.g., \cite{BPP1,BGGKP99,BGKP96,otherbb0nu,PPSNO2bb}).
We will summarize here briefly the main results 
for the three possible types of 
neutrino mass spectrum \cite{PPSNO3bb,BFS04}.
\begin{enumerate}
\item {\it Normal hierarchical neutrino mass spectrum}:
\be
m_{1}\ll m_{2}\ll m_{3}
\label{14}
\ee
For the effective Majorana mass we have 
in this case the following upper bound
\be
|m_{ee}| \lesssim \left ( \sin^{2} \theta_{12} \, \sqrt{\Delta m^{2}_{21}} +  
\sin^{2} \theta_{13}\, \sqrt{\Delta m^{2}_{32}} \right ).
\label{15}
\ee
Using the 90\% C.L. ranges (\ref{3}) and (\ref{5}) 
of the oscillation parameters and the CHOOZ bound
(\ref{6}), for the effective Majorana mass 
one finds \cite{PPSNO3bb}
\be
|m_{ee}| \lesssim 5.5\cdot 10^{-3}~\rm{eV},
\label{16}
\ee
This bound is significantly smaller than 
the expected sensitivity of the future 
$0\nu\beta\beta$-decay
experiments. The observation of 
the $0\nu \beta \beta$-decay 
in the next generation of experiments 
might exclude normal 
hierarchical neutrino mass spectrum.
\item {\it Inverted hierarchical neutrino mass spectrum}:
\be
m_{3}\ll m_{1}< m_{2}.
\label{17}
\ee
For the effective Majorana mass we have
in this case \cite{BGKP96,BGGKP99}:
\be
|m_{ee}|
\simeq \sqrt{ |\Delta m^{2}_{32}|}\,~
(1-\sin^{2} 2\,\theta_{12}\,\sin^{2}\alpha)^{\frac{1}{2}},
\label{18}
\ee
where  $\alpha = \alpha_{21}$ 
is the Majorana $CP-$ violating phase.
From eq. (\ref{18}) we obtain the range 
\be
\sqrt{ |\Delta m^{2}_{32}|}~\cos 2\,\theta_{12} \lesssim |m_{ee}|
\lesssim \sqrt{ |\Delta m^{2}_{32}|}, 
\label{19}
\ee
\noindent where the upper and 
lower bounds corresponds 
to the case of CP conservation and
the same and opposite CP-parities of neutrinos
$\nu_{1}$ and $\nu_{2}$.
Using the 90\% C.L. 
allowed values of the parameters, 
eqs. (\ref{3}) and (\ref{5}), 
for the effective Majorana mass 
one finds \cite{PPSNO3bb}:
\be
10^{-2}~\rm{eV} \lesssim |m_{ee}|\lesssim
5.5 \cdot 10^{-2}~\rm{eV}
\label{20}
\ee
Thus, if 
the neutrino mass spectrum is of the
inverted hierarchical type and the
massive neutrino 
are Majorana particles, $0\nu \beta \beta$-decay 
can be observed in the experiments of 
next generation.

\item {\it Quasi-degenerate neutrino mass spectrum}: 
\be
m_{1}\simeq m_{2}\simeq m_{3},~~ 
m^2_{1,2,3} \gg |\Delta m^2_{32}|~.
\label{21}
\ee
The effective Majorana mass 
is given in this case by eq. (\ref{18}) 
in which 
$\sqrt{ |\Delta m^{2}_{32}|}$ is replaced by
$ m_{\rm{min}}$, where
$m_{\rm{min}}$ is the lightest neutrino mass.
For the effective Majorana mass 
we have the range \cite{PPSNO3bb}
\be
0.22\, m_{\rm{min}}\lesssim |m_{ee}|\lesssim
m_{\rm{min}}.
\label{22}
\ee
In the case of quasi-degenerate neutrino mass 
spectrum, the effective Majorana mass depends 
essentially on two parameters: the lightest neutrino 
mass $m_{\rm{min}}$  and the CP-violating
parameter $\sin^{2}\alpha$. From the measurement 
of $|m_{ee}|$, the following range 
for the lightest neutrino mass
can be obtained: 
\be
|m_{ee}|\lesssim m_{\rm{min}} \lesssim 4.6\,~|m_{ee}|~.
\label{23}
\ee

  If the lightest neutrino mass $m_{\rm{min}}$
will be determined 
in the tritium $\beta-$decay experiment KATRIN 
\cite{KATRIN} which is under preparation at present,
or from cosmological and astrophysical 
observations (see, e.g., \cite{Tegmark}),
the data of $0\nu \beta \beta$ 
experiments can be used to get information 
about the Majorana CP-violating phases
\cite{BGKP96,BPP1,WRode01,PPR1}.

\end{enumerate}

   Neutrinoless double $\beta$-decay is a
unique process. The observation of this process 
would be a proof that the total lepton charge 
is not conserved and massive neutrinos 
$\nu_{i}$ are  Majorana particles. 
As we have seen in this Section, the precise 
measurement of the parameter 
$|m_{ee}|$ would allow to draw important 
conclusions about the character of neutrino 
mass spectrum, the lightest neutrino mass 
and the CP-violation associated with the 
Majorana neutrinos (for a more detailed discussion
see, e.g., \cite{PPVenice03}). However, from the data of the 
$0\nu \beta \beta$-decay experiments
only the product of $|m_{ee}|$ and the 
corresponding nuclear matrix element can be 
determined. In the next section we will briefly 
discuss the problem of the calculation of 
the nuclear matrix elements of $0\nu \beta \beta-$decay.         

\vspace{-0.4cm}

\section{The Problem of Calculation of Nuclear Matrix Elements}


  If neutrinoless double $\beta$-decay  is due 
to the Majorana neutrino mixing (\ref{1}) {\it only}, 
it proceeds via exchange of a virtual
neutrino and is a process of second order 
in the Fermi constant $G_{F}$.
The nuclear matrix element (NME) of the 
$0\nu \beta \beta$-decay of 
a given even-even nuclei 
cannot be related to other
observables and has to be calculated.
The calculation of NME is a complicated problem.
One of the problems of the calculations is 
connected with a large number of states 
of the intermediate odd-odd nuclei, which 
are important due to relatively large 
average momentum of the virtual neutrino.

  Many calculations of NME exist in 
literature (see, e.g., the review articles 
\cite{Faessler,Suhonen,Elliot,CivSuh03}). 
Two basic methods of calculations of NME 
are used at present: nuclear shell model (NSM) 
and quasiparticle random phase approximation (QRPA).
\begin{table}[ht]
\begin{center}
\caption{
Half-life of the
$0\nu \beta\beta $-decay for 
$|m_{ee}|= 5\cdot 10^{-2}\,\rm{eV}$. The nuclear matrix 
elements were taken from the compilation  
in ref. \cite{CivSuh03}.
}
\begin{tabular}{|cc|}
\hline
Nucleus
&
 $T^{0\nu}_{1/2} \rm{years}$
\\
\hline
$^{76} \rm{Ge}$
&
$1.4\cdot 10^{27}-1.5\cdot 10^{29}$
\\
$^{100} \rm{Mo}$
&
$1.7\cdot 10^{26}-5.9\cdot 10^{30}$
\\
$^{130} \rm{Te}$
&
$7.7\cdot 10^{26}-3.4\cdot 10^{27}$
\\
$^{136} \rm{Xe}$
&
$2.7\cdot 10^{27}-1.7\cdot 10^{28}$
\\
\hline
\end{tabular}
\end{center}
\end{table}

  The nuclear shell model is attractive 
from physical point of view: there are many 
spectroscopic data in favor of shell structure 
of nuclei (spins and parities of nuclei, 
binding energies of magic nuclei, etc.) \cite{ShellM1}.  
However, only rather limited set of one-particles 
states of valent nucleons can be taken into account 
because of practical computational reasons.
It is difficult to estimate the accuracy of 
the shell model calculations. 

      The most popular method of calculation 
of the NME of $0\nu\beta \beta$ -decay 
is QRPA \cite{Faessler,Suhonen}. 
This method allows to use as a basis  
a large number of one-particle states 
and to take into account all intermediate states. 
Important parameters of QRPA are the constant of 
particle-hole interaction, $g_{ph}$, and 
the constant of particle-particle 
interaction,  $g_{pp}$. 
The constant $g_{ph}$ can be fixed from a 
fit of the energy of the giant Gamov-Teller resonance. 
The constant $g_{pp}$ 
is a free parameter. 

There are many models based on the QRPA approach
(see, reviews \cite{Faessler,Suhonen,Elliot,CivSuh03}). 
The results of the calculations of NME of  
$0\nu\beta \beta$ -decay performed by different 
authors differ quite significantly.
The variety of results of the calculations 
is illustrated by Table 1 in which
ranges of the values of the half-life of the
$0\nu \beta\beta $-decay of different nuclei are presented for 
$|m_{ee}|= 5\cdot 10^{-2}\,\rm{eV}$.

Recently, in the framework of QRPA,
a new procedure of calculation of the NME
was proposed \cite{RFSV03}. 
For a fixed value of the constant $g_{ph}$
and the values of the parameter
$g_{pp}$, determined from the measured
$2\nu\beta \beta$-decay half-life,
the $0\nu\beta \beta$-decay
nuclear matrix elements of several nuclei
were calculated.
In \cite{RFSV03} results were derived
for three different sets of the one-particle 
states and for three 
different nucleon-nucleon potentials,
and  it was shown that 
the NME of $0\nu\beta \beta$-decay
depends weakly on the number of one-particle states 
and on the nucleon-nucleon potential 
used in the calculations
(the NME for each of the nuclei
thus calculated varies by not more than 10\%).

  Another approach was proposed in \cite{CivSuh03}. 
In this paper all parameters 
of the QRPA model, including the $g_{pp}$ constant,  
were fixed from data on $\beta$-decay of  
nuclei which are close to the even-even nuclei 
of interest for the $0 \nu\beta \beta$ -decay study.

  In spite of the recent progress in the calculation 
of NME, it is not possible at present to 
estimate the real accuracy of the calculations. 
It is important to find 
a possibility to check 
the calculations of NME by a 
direct comparison with experimental data \cite{BiGrif02}. 
Such a possibility will be discussed
in the next Section.

\vspace{-0.4cm}

\section{Possible Test of the NME Calculations}


   If the Majorana neutrino mixing (\ref{1}) 
is the mechanism  of $0\nu \beta\beta$-decay,  
the matrix element of the process has the 
following general form (see, e.g., \cite{Doi85,BiPet87})
\be
\langle f|(S-1)|i \rangle 
 = N\, m_{ee}\, M^{0\,\nu}(A,Z)\,\delta (E_{f}-E_{i}).
\label{25}
\ee
%
\noindent Here N is a product of known factors 
and $M^{0\,\nu}(A,Z)$ is the nuclear matrix element
of interest. The neutrino masses enter into 
the matrix element $\langle f|(S-1)|i\rangle$
through the effective Majorana mass $m_{ee}$
given by (\ref{9}) and the neutrino propagator 
$(q^{2}-m^{2}_{i})^{-1}$ which is included 
in $M^{0\,\nu}(A,Z)$, $q$ being the momentum 
of the virtual neutrino. For small neutrino masses 
(smaller than the binding energy of nucleons 
in nuclei $~\sim 10~\rm{MeV}$), the
neutrino masses in the propagator can be 
safely neglected. The matrix element 
$M^{0\,\nu}(A,Z)$ depends in this case only 
on the nuclear properties and strong interaction.

    The half-life of the $0\nu\beta\beta$ 
is given by the expression
\footnote{It follows from this expression  that
the relative accuracy of determination of the 
parameter $|m_{ee}|$ (for any value of NME) is 
two times better that the relative accuracy of 
the measurement of the half-life of 
$0\nu\beta\beta$-decay:
$$ \frac{\Delta |m_{ee}| }{|m_{ee}|}=  \frac{1 }{2}\, 
\frac{\Delta T^{0\,\nu}_{1/2} }{T^{0\,\nu}_{1/2}}$$.
}:
\be
\frac{1}{T^{0\,\nu}_{1/2}(A,Z)}=
|m_{ee}|^{2}\,|M^{0\,\nu}(A,Z)|^{2}\,G^{0\,\nu}(E_{0},Z),
\label{26}
\ee
%
\noindent where $G^{0\,\nu}(E_{0},Z)$ is known phase-space 
factor ($E_{0}$ is the energy release).
If we use a model $M$ of the calculation of NME we have 
\be
|m_{ee}|_{M}^{2}(A,Z)= \frac{1}{T^{0\,\nu}_{1/2}(A,Z)\,
|M_{M}^{0\,\nu}(A,Z)|^{2}\,G^
{0\,\nu}(E_{0},Z)}.
\label{27}
\ee

  Let us assume that neutrinoless double $\beta$-decay 
of {\em several} nuclei is observed. The effective Majorana 
mass $|m_{ee}|$ cannot depend on the parent nucleus. 
Thus, if the light Majorana neutrino exchange
is the dominant mechanism of $0\nu\beta\beta$-decay,
the model $M$ of the calculation of the nuclear 
matrix elements can be correct only  if the relations
\be
|m_{ee}|^{2}_{M}(A_1, Z_1) \simeq |m_{ee}|^{2}_{M}(A_2, Z_2)=...
\label{28}
\ee
hold, where $|m_{ee}|^{2}_{M}(A_j, Z_j)$ is the value of
$|m_{ee}|^{2}$ obtained from the $0\nu\beta\beta$-decay
half-life of the nucleus $(A_j, Z_j)$ using the model
$M$.

  Consider different models of calculation of NME.
From eq. (\ref{26}) it follows 
that for a given parent nucleus the product 
$|m_{ee}|_{M}^{2}(A,Z)|\,|M_{M}^{0\,\nu}(A,Z)|^{2}$
does not depend on the model.
Thus, for different models and the same nucleus we have
\be
|m_{ee}|_{M_1}^{2}(A,Z)|\,|M_{M_1}^{0\,\nu}(A,Z)|^{2}=
|m_{ee}|_{M_2}^{2}(A,Z)|\,|M_{M_2}^{0\,\nu}(A,Z)|^{2}=...
\label{29}
\ee

\begin{table}[ht]
\begin{center}
\caption{
The parameter $\eta^{M_{i};M_{k}}(A, Z)$, determined by
eq. (\ref{31}), for
the nuclear matrix elements of 
$0 \nu \beta \beta$- decay, calculated in 
ref. \cite{ShellM99} ($M_{1}$), in 
ref. \cite{RFSV03} ($M_{2}$) and in ref. \cite{CivSuh03} ($M_{3}$).
}
\vspace{0.4cm}
\begin{tabular}{|cccc|}
\hline
Nucleus
&
 $\eta^{M_{2};M_{1}}$
&
$\eta^{M_{3};M_{1}}$
&
$\eta^{M_{2};M_{3}}$
\\
\hline
$^{76} \rm{Ge}$
&
0.37
&
0.19
&
1.93
\\
$^{82} \rm{Se}$
&
---
&
0.38
&
---
\\
$^{100} \rm{Mo}$
&
---
&
---
&
6.56

\\
$^{130} \rm{Te}$
&
0.74
&
0.10
&
7.32
\\
$^{136} \rm{Xe}$
&
0.53
&
0.02
&
22.42
\\
\hline
\end{tabular}
\end{center}
\end{table}
\noindent For two different models we have the relation
\be
|m_{ee}|_{M_2}^{2}(A, Z)  =\eta^{M_{2};M_{1}}(A, Z)\, |m_{ee}|_{M_2}^{2}(A, Z) 
\label{30}
\ee
where
\be
\eta^{M_{2};M_{1}}(A, Z) = \frac{ |M^{0\,\nu}_{M_1}(A,Z)|^{2}} 
{|M^{0\,\nu}_{M_2}(A,Z)|^{2}}~.
\label{31}
\ee

   In Table 2 we present the values of the coefficient 
$\eta(A, Z)$ for the case
of the  matrix elements calculated in 
\cite{ShellM99} (NSM), 
in \cite{RFSV03} (QRPA ), and 
in \cite{CivSuh03} (QRPA, different model).  

  We see from  Table 2 that the coefficient $\eta(A,Z)$ 
depends rather strongly on $(A,Z)$. This means that if 
for one model the relation (\ref{28}) is satisfied, 
other models, in principle, can be excluded. 
However, the observation of $0\nu\beta\beta$-decay of 
only two nuclei might not allow 
to distinguish between different models. For example, the observation 
of the $0\nu\beta\beta$-decay of $^{100} \rm{Mo}$ and $^{130} \rm{Te}$ 
will not allow to distinguish the QRPA models of ref. \cite{RFSV03} 
and of ref. \cite{CivSuh03} (the difference between 
the values of the  coefficient $\eta(A, Z)$ for 
these two nuclei is about 10\%). 
The values of the effective Majorana mass
which can be obtained from the observation of 
the $0\nu\beta\beta$-decay of these two nuclei, 
if one uses the models of ref. \cite{RFSV03} and
of ref. \cite{CivSuh03}, will
differ by a factor of $\sim 2.6$. 
The observation of 
neutrinoless double $\beta$-decay of at least 
three nuclei would  be an important 
tool in the solution of the problem of NME.
Table 2 suggests that 
the observation of the $0\nu\beta\beta$- decay of $^{76} \rm{Ge}$,
$^{130} \rm{Te}$ 
and $^{136} \rm{Xe}$ 
would solve the problem..
 
 If relations (\ref{28}) are satisfied for some model $M$,
this would mean that the corresponding value 
of the effective Majorana mass $|m_{ee}|_{M}$ can 
be different from the true value $|m_{ee}|_{0}$ 
by a constant factor:
\be
|m_{ee}|_{M}=\beta\, |m_{ee}|_{0}.
\label{32}
\ee
From this relation it follows that
\be
|M_{M}^{0\,\nu}(A,Z)|_{M}=\frac{1}{\beta}\,|M_{M}^{0\,\nu}(A,Z)|_{0},
\label{33}
\ee
where $|M_{M}^{0\,\nu}(A,Z)|_{0}$ is the true 
value of the NME (which, of course, we do not know). 
For the test we are proposing 
it is important that nuclear matrix 
elements depend rather strongly on $(A,Z)$ . 

It looks quite improbable 
that relation (\ref{33}) 
with one and the same constant  $\beta$
would be valid for nuclei 
with different properties 
and different NME, and  
especially for three different nuclei. 
Thus, if  relation  (\ref{28}) 
is satisfied 
for some model of calculations of nuclear
matrix elements,
the corresponding value 
of the effective Majorana mass would 
most likely be the true value.

  One last remark. We have assumed that the 
mechanism of $0\nu\beta\beta$-decay is 
the Majorana neutrino mixing, eq. (\ref{1}). 
There exist, however,
many other mechanisms of nonconservation of the 
total lepton charge and $0\nu\beta\beta$-decay, 
like SUSY with violation of $R-$parity, etc. 
(see, e.g., the review
articles \cite{Faessler, Suhonen}).
These additional mechanisms modify 
the neutron-proton-electron 
vertexes and include exchange
not only of light neutrino but also of 
heavy particles (see, e.g., \cite{Pas}). 
Because the nuclear matrix elements of different 
operators  have different $A,Z$ dependence, 
in the case of additional mechanisms 
the factorization property of 
$0\nu\beta\beta-$decay matrix elements, 
eq. (\ref{25}), 
will not be valid and relation (\ref{28}), in 
general, will not be satisfied. 
Thus, if relation (\ref{28}) 
will be found to hold for a given model,
this also will be a strong 
indication in favor of dominance 
of the Majorana neutrino mixing 
mechanism of lepton charge 
non-conservation.

\vspace{-0.4cm}

\section{Conclusions}

   
  After the discovery of  neutrino masses 
 and neutrino mixing  the problem
of {\em the nature of neutrinos with definite masses $\nu_{i}$} 
became one of the fundamental problems in the 
studies of neutrino mixing.
The determination of the nature of
massive neutrinos $\nu_{i}$ 
will have a profound 
implications for the understanding of
the mechanism of generation of 
neutrino masses and mixing. 
The measurement of the 
effective Majorana mass would allow
to obtain an important and unique information about 
the character of the neutrino mass spectrum, 
the lightest neutrino mass and possibly on the
Majorana CP-violating phases.

      From the measured half-life of  $0\nu\beta\beta$-decay
only the product of the effective Majorana mass and
the corresponding nuclear matrix element can be obtained.
The results of the  different existing calculations of the
$0\nu\beta\beta$-decay nuclear matrix elements
vary significantly (see Table 1). The improvement of 
the calculations of the nuclear matrix 
elements is a very important and challenging problem.
We have discussed here a possible method which 
could allow to test the models of calculation of NME of the
$0\nu\beta\beta$-decay via comparison of the 
results of calculations with experimental data.
The method is based on the factorization 
property of the matrix element of $0\nu\beta\beta$-decay 
and requires the observation of the 
$0\nu\beta\beta$-decay of several nuclei.

  The  nuclear matrix elements  
of the $0\nu\beta\beta$-decay cannot be 
related to other observables. 
We do not see at present any alternative 
possibility to confront the results 
of the NME calculations with experimental data in 
a model independent way. 

    New experiments searching for $0\nu\beta\beta$-decay
of $^{130} \rm{Te}$, $^{76} \rm{Ge}$, 
$^{136} \rm{Xe}$,  $^{100} \rm{Mo}$ and other nuclei are in
preparation at present. From the point of view of 
the problem of NME, it is very desirable 
that these projects will be realized  
{\it for at least three different nuclei}. 
{\it The selection of the nuclei
should be done taking into account 
also the considerations discussed above
regarding the possibility to test
the nuclear matrix element calculations}.

\vspace{-0.4cm}
\section{Acknowledgments}
\vspace{-0.3cm}
   We are grateful  to F.\v Simkovic for providing us with 
detailed information about the results 
obtained with the method proposed in ref. \cite{RFSV03}.
This work was supported in part by
the Italian MIUR and INFN under the programs
``Rientro dei cervelli'' (S.M.B.) and 
``Fisica Astroparticellare''(S.T.P.).
S.T.P. would like to thank  
Prof. T. Kugo, Prof. M. Nojiri and 
the other members of the Yukawa Institute 
for Theoretical Physics (YITP), Kyoto, Japan,
where part of the work on this article was done,
for the kind hospitality extended to him.

\end{document}